# Classification of Random Boolean Networks

Carlos Gershenson[†,‡]
[†]School of Cognitive and Computer Sciences
University of Sussex
Brighton, BN1 9QN, U. K.
C.Gershenson@sussex.ac.uk
http://www.cogs.sussex.ac.uk/users/carlos
[‡]Center Leo Apostel, Vrije Universiteit Brussel
Krijgskundestraat 33, B-1160 Brussels, Belgium

**Abstract**

We provide the first classification of different types of Random Boolean Networks (RBNs). We study the differences of RBNs depending on the degree of synchronicity and determinism of their updating scheme. For doing so, we first define three new types of RBNs. We note some similarities and differences between different types of RBNs with the aid of a public software laboratory we developed. Particularly, we find that the point attractors are independent of the updating scheme, and that RBNs are more different depending on their determinism or non-determinism rather than depending on their synchronicity or asynchronicity. We also show a way of mapping non-synchronous deterministic RBNs into synchronous RBNs. Our results are important for justifying the use of specific types of RBNs for modelling natural phenomena.

## 1. Introduction

Random Boolean Networks (RBNs) have been used in diverse areas to model complex systems. There has been a debate on how suitable models they are depending on their properties, mainly on their updating scheme (Harvey and Bossomaier, 1997; Di Paolo, 2001). Here we make a classification of different types of RBNs, depending on their updating scheme (synchronous, asynchronous, deterministic, non-deterministic), in order to study the properties, differences, and similarities of each one of them. The aim of this study is to increase the criteria for judging which types of RBNs are suitable for modelling particular complex systems.

In the next section we present our classification of RBNs, first mentioning the previously defined types of RBNs, and then defining three new types of RBNs. In Section 3 we present the results of a series of experiments carried out in a software laboratory, developed by us specially for this purpose, and available to the public. Particularly, we present our studies on point attractors, statistics of attractor density in different types of network, and the homogeneity of RBNs (which validates our statistical analyses). In Section 4 we propose a (non-optimal but general) method for mapping deterministic RBNs into synchronous RBNs. In Section 5 we identify further directions for studying and understanding different types of RBNs. We conclude estimating the value of our work, and we use our results to clarify a controversy on the proper use of RBNs as models of genetic regulatory networks (Kauffman, 1993; Harvey and Bossomaier, 1997).

## 2. Random Boolean Networks

There has been no previous classification of different types of RBNs, perhaps because of the novelty of the concept, but there has been enough research to allow us to make a classification of RBNs according to their updating schemes.

Classical Random Boolean Networks (CRBNs) were proposed by Stuart Kauffman (1969) to model genetic regulation in cells. They can be seen as a generalization of (binary) cellular automata (CA). These were first developed in the late 1940's by John von Neumann (1966), for studying self-replicating mechanisms.

A RBN has n nodes, which have connections (k) between them (n, k $\in \mathbb{N}$, k$\leq$n). The state (zero or one) of a node at time t+1 depends on the states at time t of the nodes which are connected to the node. Here we will limit our study only to homogeneous RBNs, where k is the same for all nodes. Logic rules (n*$2^k$ values in lookup tables) are generated randomly, and they determine how the nodes will affect each other.

In CRBNs, the updating of the nodes is synchronous, in the sense that all the nodes in the network are updated at t+1 depending their states at t. Boolean random maps are equivalent to CRBNs with n=k. The dynamics arising from CRBNs are very interesting, since depending on the values of n and k, they can be ordered, complex, or chaotic (Kauffman, 1993; Wuensche 1998). Therefore, they have been also used to study these types of dynamics in deterministic systems. Diverse properties of CA and CRBNs have been studied, among others, by Wuensche (1997; 1998) and by Aldana, Coppersmith and Kadanoff (2002). They have been used to model phenomena in physics, biology, cognition, economy, sociology, computation, artificial life, and complex systems in general (Wuensche, 1998).

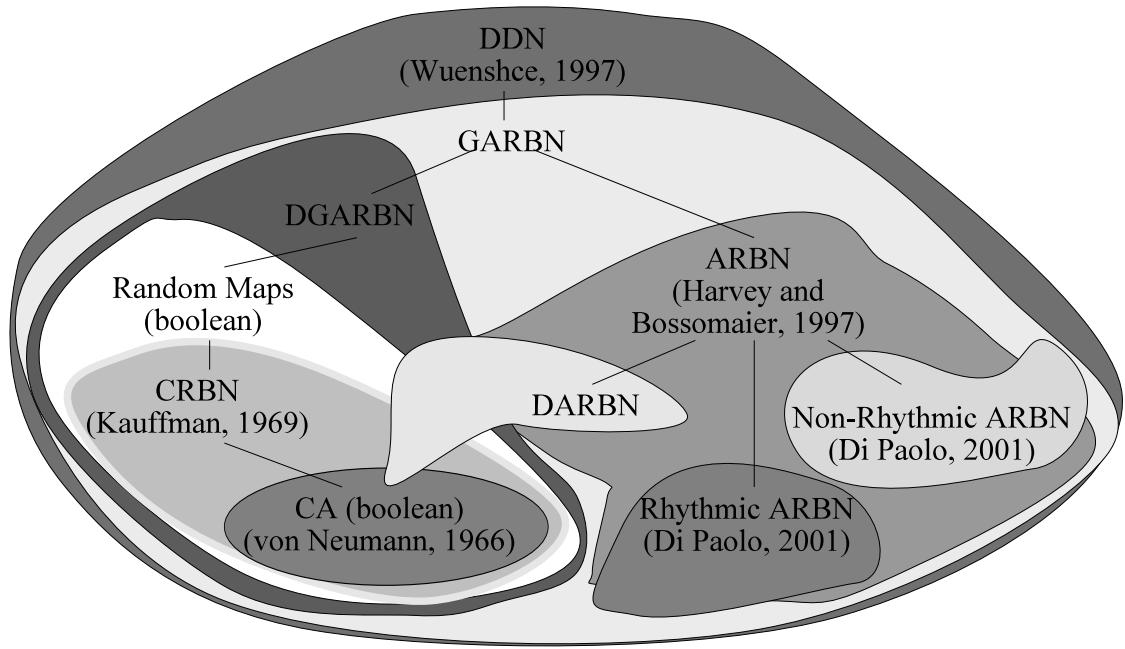

Figure 1. Classification of Random Boolean Networks

CRBNs have attractors, which consist of a state ("point attractor") or a group of states ("cycle attractor") which "capture" the dynamics of the network. Since CRBNs are deterministic, once the state of a network reaches an attractor, it will never have states different from the ones in the attractor. And since the number of states of a CRBN is also finite, in theory, an attractor will always be reached.

Harvey and Bossomaier (1997) studied Asynchronous Random Boolean Networks (ARBNs) They have the characteristics of CRBNs, but their updating is asynchronous. Well, in ARBNs, the updating is not only asynchronous, but also random. Each time step a single node is selected at random in order to be updated. Because of this, in ARBNs there are no cycle attractors (although there are point attractors). The system cannot escape from a point attractor, because no matter which node is selected, the state of the network will not change. ARBNs also have "loose attractors" (Harvey and Bossomaier, 1997), which are parts of the state space which also capture the dynamics, but since the updating order of the nodes is random, the order of the states will not be repeated deterministically. Although, Di Paolo (2001) evolved successfully ARBNs for finding rhythmic and non-rhythmic attractors, but ARBNs with these attractors seem to be a very small subset of all possible ARBNs.

In order to have a more complete taxonomy of RBNs, we define three types of RBNs, setting all of them and the previously exposed under Discrete Dynamic Networks (DDNs), a term introduced by Wuensche[1] (1997), since we can say that DDNs have all the properties common in all types of RBNs: they have discrete time, space, and values. DDNs outside of our scope would be multivalued networks, but they would be DDNs also. Real-valued networks would not be DDNs, since their values are not discrete, but continuous. Dynamical Systems Theory studies such systems.

We define Deterministic Asynchronous Random Boolean Networks (DARBNs) as ARBNs which do not select at random which node to update. Each node has associated two parameters: p and q (p, q $\in \mathbb{N}$, q<p), which determine the period of the update (how many time steps the node will wait in order to be updated), and the translation of the update, respectively. A node will be updated when the modulus of time t over p is equal to q. If two or more nodes will be updated at a specific time, the first node is updated, and then the second is updated taking into account the new state of the network. This order is arbitrary, but since there is no restriction in the connection of the nodes by their locality (as with CA), we can have any kind of possible DARBNs with this updating scheme. The advantage is that with DARBNs we can model asynchronous phenomena which are not random, a thing which is quite difficult with ARBNs. Therefore, DARBNs have cycle (and point) attractors.

ARBNs have another limitation: they only update one node at a time. We define Generalized Asynchronous Random Boolean Networks (GARBNs) as ARBNs which can update any number of nodes, picked at random, at each time step. This means that GARBNs can go from not updating any node at a time step, passing to updating only one (as ARBNs), updating some nodes synchronously, to updating all the nodes synchronously (as CRBNs). As ARBNs, they are non-deterministic, so again there are no cycle attractors, only point and loose attractors.

---

[1]Although he studied only synchronic DDNs in (1997), the terminology is appropriate.

As we did with DARBNs, we introduce the parameters p and q associated to each node to define Deterministic Generalized Asynchronous Random Boolean Networks (DGARBNs). They do not have the arbitrary restriction of DARBNs, so if two or more nodes are determined by their p's to be updated at the same time step, they will be updated synchronously, *i.e.* they will all be updated at time t+1 taking into account the state of the network at time t. Note that DGARBNs and DARBNs overlap in the specific cases when p and q are such that one and only one node is updated each time step (for example, a network of two nodes (n=2), one being updated at even time steps (p=2, q=0), and another being updated at uneven time steps (p=2, q=1)).

Also other specific configurations produce the same behaviour for all the types of network independently of their updating schemes (*e.g.* when k=0, or when all states are point attractors).

Figure 1 shows a graphic representation of the classification just described above.

As we stated, we classify all RBNs under DDNs. The most general are GARBNs, since all the others are particular cases of them. If on one hand, we make them deterministic with parameters p and q, we will have DGARBNs. Random maps are emerging when n=k, and for all nodes p=1 and q=0. These ones can simulate with redundancy any CRBN, but not otherwise. Therefore, CRBNs can be seen as a subset of random maps. Boolean CA are specific cases of CRBNs, where the connectivity is limited by the spatial organization of the nodes. On the other hand, if we limit GARBNs for updating only one node at a time, we will have ARBNs. If we make them deterministic, we will have DARBNs. There are also special cases of ARBNs with rhythmic and non-rhythmic attractors. Very probably GARBNs with rhythmic and non-rhythmic attractors can be found. Most types of RBNs can behave in the same way in limit cases (*e.g.* when k equals zero, or when the number of attractors equals n).

| RBN | updating scheme |
|---|---|
| CRBNs | synchronous, deterministic |
| ARBNs | asynchronous, non-deterministic |
| DARBNs | asynchronous, deterministic |
| GARBNs | semi-synchronous, non-deterministic |
| DGARBNs | semi-synchronous, deterministic |

Table 1. Updating schemes of RBNs.

Here we will study the properties of CRBNs, ARBNs, DARBNs, GARBNs, and DGARBNs. Table 1 shows the characteristics of the updating schemes for each one of these RBNs. Note that there are no synchronous, non-deterministic RBNs. We say that GARBNs and DGARBNs have semi-synchronous updating because in some cases they can behave synchronously (all nodes updated at once), and in some cases as asynchronously (only one node is updated at once), but mainly in all the possibilities in between, *i.e.* when *some* nodes are updated synchronously.

As a simple example, in Table 2 we show the transition table of a RBN of n=2, k=2, p's={1,2}, and q's={0,0}. Figure 2 shows the different trajectories that the RBN will have, depending on its updating scheme.

| net(t) | net(t+1) |
|---|---|
| 11 | 11 |
| 10 | 01 |
| 01 | 00 |
| 00 | 10 |

Table 2. Transition table for a RBN n=2, k=2.

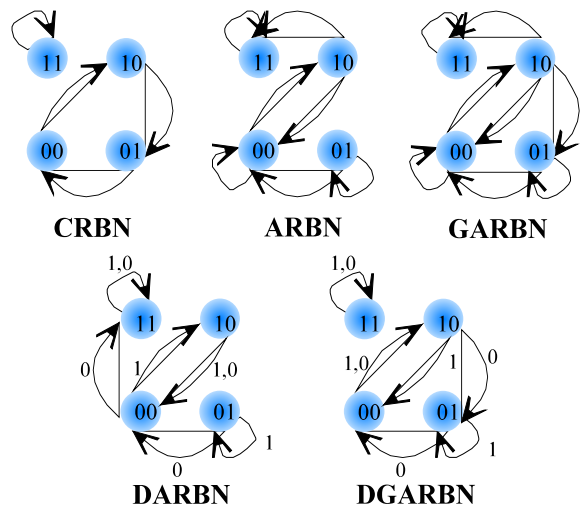

Figure 2. Graphs of RBNs with different updating schemes. The arrows with numbers indicate which transition will take place depending on the modulus of time over two.

We can see that all types of updating schemes produce different behaviours of the RBN. Some of them are more similar than others, though. In the next section we explore the properties of different types of RBNs.

# 3. Experiments and Analysis

We developed a software laboratory for studying the properties of different types of RBNs. It is available online to the public (Java source code included) at http://www.cogs.sussex.ac.uk/users/carlos/rbn. It can be used through a web browser, providing a friendly interface for generating, editing, displaying, saving, and loading RBNs, and analysing their properties. The results presented in this section are product of experiments carried out in our laboratory.

## 3.1. Point Attractors

Our first finding is that the point attractors are the same for any type of RBN. In other words, if we find a point attractor in an ARBN, if we change the updating scheme to CRBN (or any other), the point attractor will be the same. This is because a point attractor is given when for all the nodes, their rules determine that when they will be updated, they will have the same value. Therefore, it is not important in which order or how many nodes are updated, since none will produce a change in the state of the network.

We should note that the basins of the point attractors in most cases are very different for different types of RBNs. The cycle and loose attractors and their basins also change (tough not always) depending on the updating scheme. But for a determined network connectivity and rules, the point attractors will be the same.

## 3.2. Attractor density

For this section, we obtained with our laboratory statistics from 1000 networks of each of the presented values of n and k. First we generated randomly a RBN with the specified n, k, p, and q, and then we tested the RBN with the different updating schemes. For each type of network we tested all possible initial states, running them for 10000 time steps, expecting to reach an attractor. Then we searched for attractors of period smaller than 50 for CRBNs and 200 for the other deterministic cases (checking that the state and periods would be the same as the ones as t=10000), and point attractors for the random cases (if state(10000)=state(10001)=...=state(10050)). For all our experiments the p's for all the nodes were generated randomly, taking values between 1 and 4, and all q's=0.

Figure 3 shows the average number of deterministic attractors for networks of n's between 1 and 5, with all their possible k's (from 0 to n). Figure 4 shows the same information for networks of k=3 and n between 3 and 8. Remember that RBNs which are updated with random schemes (ARBNs and GARBNs) can only have deterministic attractors of period one (point attractors), and if one network has a point attractor, the other types of network will have the same point attractor. We can see from Figure 3 that the average of point attractors in ARBNs is roughly lower for k=3, consistent with results of Harvey and Bossomaier (1997)[2]. From both figures we can see that DARBNs, and DGARBNs even more, are very close to the number of attractors found in CRBNs. We can see here that there is a very big difference due to the randomness of the updating, and not so much due to the degree of synchronicity.

We can also see that, for deterministic RBNs, the growth of the average number of attractors as we increase n, with constant k, seems to be linear, consistent with the results of Bilke and Sjunnesson (2002) for CRBNs.

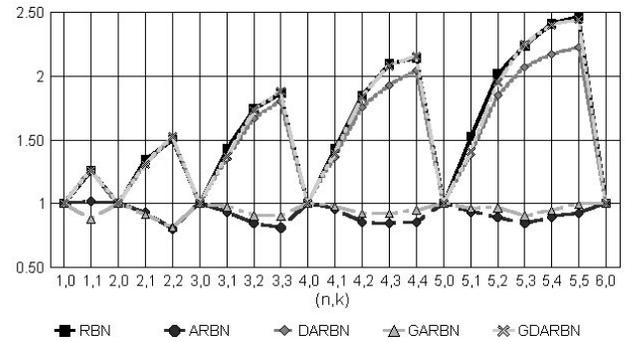
Figure 3. Average number of deterministic attractors.

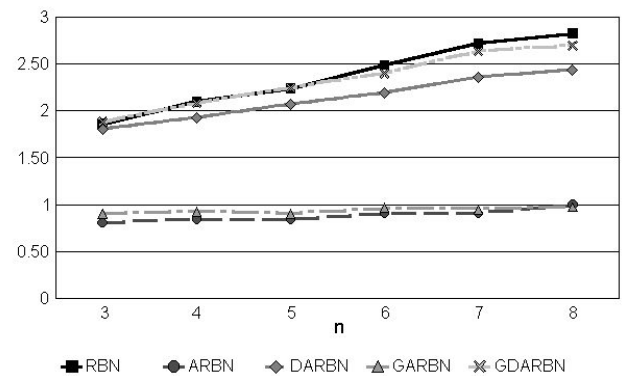
Figure 4. Average number of deterministic attractors, k=3.

Note that when there is no connectivity (k=0), there is no interesting updating, since the networks, independently of their initial state or updating scheme, will reach a final state.

---

[2] Even when the *expected* number of point attractors is one. This is, mathematically we can find that if we consider all possible ARBNs, their number is equal to the possible point attractors (Harvey and Bossomaier, 1997). Of course some networks will have no point attractors, or more than one, but if our search would be exhaustive, the average would be one. Since our search is not exhaustive, what we can see is that for k=3, the point attractors "hide themselves" more than for other values of k (there are few networks with several point attractors and many networks with none).

It is because of this that all RBNs have one and only one point attractor for any value of n when k=0.

We can appreciate the percentage of states which belong to an attractor for the same networks in Figure 5 and Figure 6. Here we can see that RBNs have more states in attractors than DGARBNs and DARBNs. This suggests that the average period of the cycle attractors is higher for RBNs.

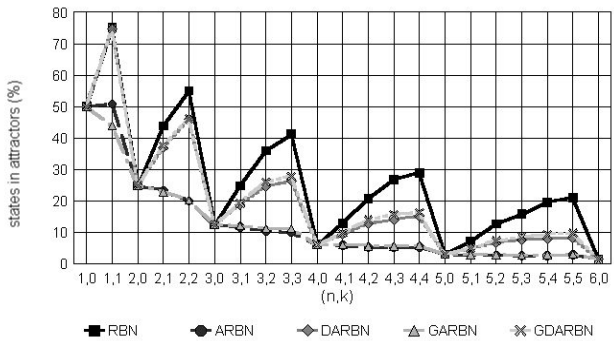

Figure 5. Percentage of states in attractors.

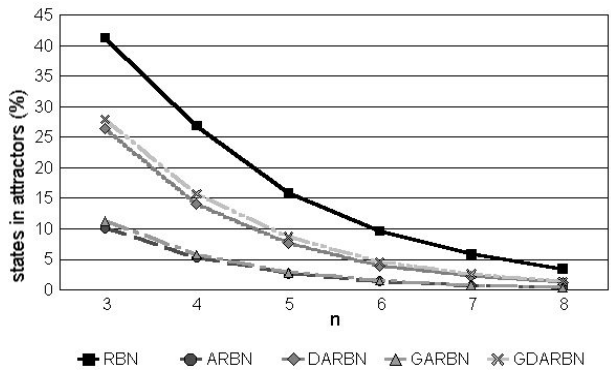

Figure 6. Percentage of states in attractors.

We can also see in Figure 5 that for deterministic RBNs, the number of states in an attractor increases with k. But the percentage of the states in attractors seems to decrease exponentially as n increases for all types of RBN.

We would also like to know exactly how many states there are in each attractor, not only their percentage. But we should note that, even when CRBNs, ARBNs, and GARBNs have $2^n$ possible states, this is increased in DARBNs and DGARBNs, because of the introduction of the updating periods. This is, a network with the same state, may be updated in a different way depending on the time it reaches that state. Therefore, in DARBNs and DGARBNs we need to take into account the least common multiple of all p's, and we will have $LCM(p_i) * 2^n$ possible states. As we can see, the periods of the attractors in DARBNs and DGARBNs grow in comparison to CRBNs, but since there are also more states, the ratio between states in attractors and total states is equivalent. But for calculating the number of states, we normalized the number of states dividing them by their total number of states divided by $2^n$.

Note that this produces a change only in DARBNs and DGARBNs (since all other RBNs have $2^n$ states), setting them in values comparable to the ones in the other RBNs. These results can be seen in Figure 7 and Figure 8.

The values of ARBNs and GARBNs are the same as the ones in Figure 3 and Figure 4, since the states in attractors are point attractors. Again we can see that the period of the cycle attractors are higher for RBNs than for DGARBNs, and these are a bit higher than the ones of DARBNs. In deterministic RBNs we can see that the number of states increases with k. They also seem to increase with n, but CRBNs increase this number much faster than the normalized DGARBNs and DARBNs. This increment for the deterministic networks (as we increase n) appears to be linear. The steepness of this increment seems to be increased with k. But as we have seen, the percentage of states in deterministic attractors decreases with n, since the number of states of the network is doubled for each node we add to the network.

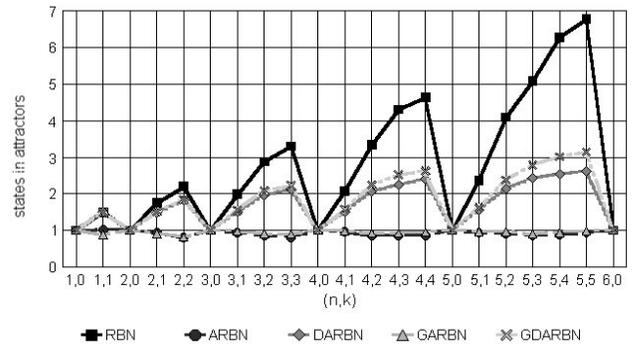

Figure 7. States in attractors (relative).

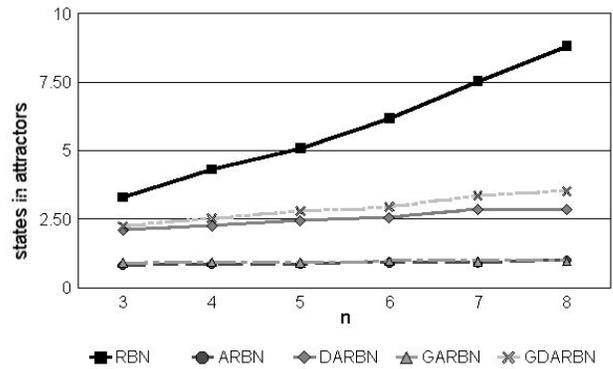

Figure 8. States in attractors (relative).

Finally, we obtained the percentage of the initial states which did not reach an attractor (after more than 10000 time steps). The results can be appreciated in Figure 9 and Figure 10. All CRBNs reached an attractor, and very few DARBNs and DGARBNs had attractors of period larger than 200 (less than 0.1%, only in large networks), so the interesting results are only for ARBNs and GARBNs. This percentage reflects the size of the basins of attraction of their point attractors

(that is, the states which **do** reach an attractor), and of course the size of the basins of attraction of the loose attractors. It seems, but is not clear, that the percentage of states which do not reach point attractors increases as n does. On the other hand, it is obvious that there is a lower probability to reach a point attractor when point attractors are harder to find (k=3).

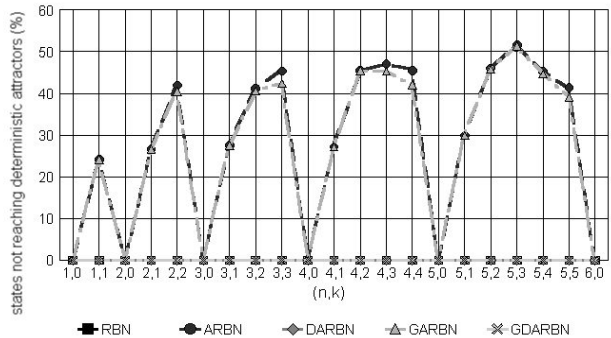

Figure 9. Percentage of initial states not reaching a deterministic attractor.

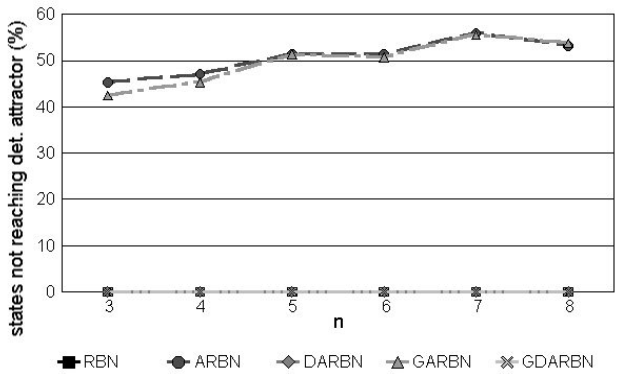

Figure 10. Percentage of states not reaching a deterministic attractor.

From the information gathered in all these charts, we can see that there is a regular order (with very few exceptions) on the number of attractors, and states in attractors for the different types of RBNs. CRBNs are on top, having the highest number of attractors and states in attractors. On the bottom, there are ARBNs, which have the lowest values. In the middle of the rest, there are DARBNs, and the generalized versions are in the spaces left in between, but very close and above DGARBNs of DARBNs and GARBNs of ARBNs.

The reason of this can be explained with the different updating schemes. RBNs with non-deterministic updating will have less deterministic attractors because they can only have point attractors. Anyway, GARBNs have a slightly higher probability than ARBNs to reach one faster since their updating is semi-synchronous, and this should enlarge their basins of attraction. Since RBNs with deterministic updating have cycle attractors, this explains why they have more attractors and states in attractors than non-deterministic ones. But we can see that these numbers increase from asynchronous to semi-synchronous to synchronous. The lack of synchronicity increases the complexity of the RBN, because we need parameters p and q to make the updating, enhancing the number of possible states and interactions. And this complexity changes the attractor basins, transforming and enlarging them. This reduces the number of attractors and states in attractors.

### 3.3. Possible Networks and Statistics

We saw that for a specific network with homogeneous connectivity, there are $n*2^k$ (binary) values in the rules of the network. Therefore, there are $2^{n*2^k}$ possible networks.

For any fixed n and k, if we change the connectivity, we will have networks redundant with these $2^{n*2^k}$ (in the number of attractors, attractor basins, etc.). Thus, if we are not interested in the particular connectivity, but on the general properties of RBNs, we will find "only" $2^{n*2^k}$ equivalent networks. To get a broad picture of how fast the number of possible networks grow, Table 3 shows the number of possible networks for small n's and k's.

| n \ k | 0 | 1 | 2 | 3 | 4 | 5 | 6 |
|---|---|---|---|---|---|---|---|
| 1 | 2 | 4 | | | | | |
| 2 | 4 | 16 | 256 | | | | |
| 3 | 8 | 64 | 4096 | 1.68E+07 | | | |
| 4 | 16 | 256 | 65536 | 4.29E+09 | 1.84E+19 | | |
| 5 | 32 | 1024 | 1048576 | 1.10E+12 | 1.21E+24 | 1.46E+48 | |
| 6 | 64 | 4096 | 16777216 | 2.81E+14 | 7.92E+28 | 6.28E+57 | 3.94E+115 |
| 7 | 128 | 16384 | 2.68E+08 | 7.21E+16 | 5.19E+33 | 2.70E+67 | 7.27E+134 |
| 8 | 256 | 65536 | 4.29E+09 | 1.84E+19 | 3.40E+38 | 1.16E+77 | 1.34E+154 |
| 9 | 512 | 262144 | 6.87E+10 | 4.72E+21 | 2.23E+43 | 4.97E+86 | 2.47E+173 |
| 10 | 1024 | 1048576 | 1.10E+12 | 1.21E+24 | 1.46E+48 | 2.14E+96 | 4.56E+192 |

Table 3. Possible equivalent networks for n nodes and k connections ($2^{\wedge}(n*(2^{\wedge}k))$).

As we can see, the number of possible networks grows tremendously fast. So, if we managed a statistical space of only 1000 sample networks, how representative will it be for n=8, k=3, if our sample is roughly $1.84*10^{-16}$ of the possible networks?

For testing how diverse our different samples were, we compared several samples of the same values of n and k, in order to see how much do they diverge. Particularly, we made five samples of 1000 networks with n=4, k=4, where the number of possible networks is $1.84*10^{19}$. For the average number of attractors, we calculated the percentage of the differences between the maximum and minimum values, and the maximum and the average values, obtained in the samples. The results can be appreciated in Figure 11. Note that the difference between the average and the minimum is the difference between the two calculated differences. Therefore, the sample space is a bit skewed. Anyway, we could see (at least for this case) that the farthest we can get from an average value is less than 5%, even when our samples are taking $1.84*10^{-16}$ of the possible networks. This means that the global properties of RBNs are very uniform. This is, it is not that there are no networks with n attractors, but there is only one. There are also very few networks with all their states in an attractor, and there are many networks with the characteristics we extracted in our experiments. All the values that we extracted in different trials in our experiments also were very close to each other.

Also, since the plots we obtained for attractor density (Figures 3-8) fir very well with linear or exponential curves, it appears that our statistics match the actual attractor density without much error. This is not the case for Figures 9-10.

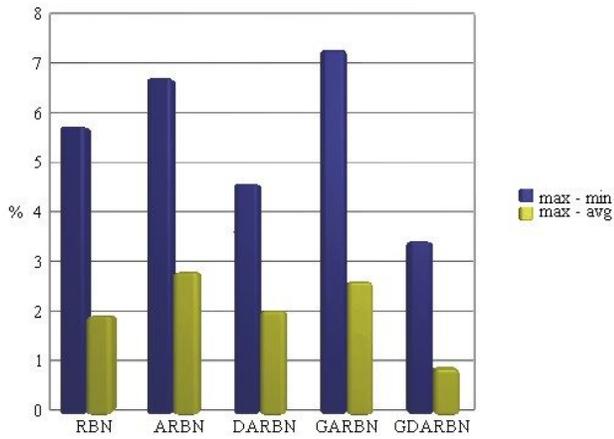

Figure 11. Percentages of differences between five samples of 1000 networks (number of attractors).

## 4. Mapping Deterministic RBNs to CRBNs

If a RBN is deterministic, we can map it to a CRBN. Of course, this is interesting only if the RBN is a DARBN or a DGARBN.

Perhaps this is not optimal, but a way of obtaining the same behaviour of a DARBN or DGARBN of specific n, k, p's, and q's is to create a CRBN of n+m nodes and k+m connectivity, where $2^m \geq LCM(p_i)$. We use the least common multiple of all p's in order to contemplate all the possible combinations of different nodes being updated, but in specific networks this might be redundant. The m nodes to be added should encode in binary base the time modulus $LCM(p_i)$. What was a function of time in the DARBN or DGARBN, now is a function in the network. Of course there is a redundancy at least when in the m nodes encoding time there is a binary value greater than $LCM(p_i)$.

In Table 4 we can appreciate the mapping to a CRBN of the DARBN with transitions shown in Table 2 and graph in Figure 2. Table 5 shows the mapping for the case of the respective DGARBN. The node m ($2^m \geq LCM(p_i)$, $LCM(p_i)=2$, m=1) which is added is shown in grey. These CRBNs have the same behaviour than the deterministic non-synchronous networks from Figure 2, but the network itself is more complex (n=3, k=3).

It would be too rushed to say that therefore all deterministic non-synchronous RBNs can be seen as CRBNs, since we do not know how redundant the proposed mapping is, and therefore the complexity of the network might be increased too much. It should still be studied how similar are the properties of CRBNs and deterministic RBNs mapped to CRBNs.

We can see that the fact that CRBNs have a higher attractor density is related to the fact that DARBNs and DGARBNs are more complex than a CRBN of the same n and k (because of the p's and q's). As we have seen in our experiments, the attractor density of RBNs decreases exponentially as we increase n. If we want to map the behaviour of a deterministic non-synchronous RBN to a CRBN, we need to add m nodes ($2^m \geq LCM(p_i)$), and such CRBN would see its attractor density reduced, just as DARBNs and DGARBNs have a lower attractor density.

| net(t) | net(t+1) |
|--------|----------|
| 111    | 110      |
| 101    | 000      |
| 011    | 010      |
| 001    | 100      |
| 110    | 111      |
| 100    | 001      |
| 010    | 001      |
| 000    | 111      |

Table 4. Map of DARBN to CRBN.

| net(t) | net(t+1) |
|--------|----------|
| 111    | 110      |
| 101    | 000      |
| 011    | 010      |
| 001    | 100      |
| 110    | 111      |
| 100    | 011      |
| 010    | 001      |
| 000    | 101      |

Table 5. Map of DGARBN to CRBN.

## 5. Unattended Issues

In order to understand better the differences of the different types of RBNs, we should also study how similar or different are the attractor basins of the RBNs as we change the updating schemes. From our experience, we have seen that sometimes they can be very similar, and sometimes very different, but we do not have yet any clear criteria for determining when each of these takes place. Also, for non-deterministic RBNs, different attractor basins might overlap (but not the attractors!). Tools such as DDLab (Wuensche, 1997; 1998) have proven to be very useful for studying attractor basins and many other properties of CRBNs and CA. Similar tools should be developed for studying further properties of different types of RBNs, such as garden-of-Eden densities. Also, to our knowledge, the basins of loose attractors are an area virgin for formal study. The study of order-complexity-chaos in the types of RBNs defined here also needs to be addressed. The study of non-homogeneous RBNs and hybrid RBNs should also increase our understanding of these models. Finally, since cellular automata could be seen as special cases of RBNs, the results presented here can be applied for understanding the differences caused by the updating schemes of cellular automata.

Because of the computational complexity of RBNs, it is difficult to study their mathematical properties straightaway. We believe this is a challenge for mathematicians, which could be addressed with the aid of our software laboratory. For example, it would be very useful to find formulae for determining the information we obtained statistically for any given n and k, especially for non-deterministic RBNs, where statistics seem to be more evasive. This would also validate or nullify the results presented here.

All these issues should be addressed if we desire to increase our understanding of RBNs.

## 6. Conclusions

In this work we presented the first proposed classification of different types of random boolean networks, depending on the synchronicity and determinism of their updating schemes. While doing so, we defined three new types of RBNs: DARBNs, GARBNs, and DGARBNs. Using a software laboratory we developed (source code available), we obtained some of the general properties of the different RBNs through statistical analysis, noticing when they have similarities and differences, but further study is required to fully understand all the properties of different RBNs.

CRBNs and ARBNs are different mostly because the first ones are deterministic and the second are not, but CRBNs are not that different to non-synchronous deterministic RBNs. DARBNs are much more similar to CRBNs than to ARBNs. We agree to the critics to synchronous CRBNs in the sense that when they are used to model any phenomena, their synchronicity needs to be justified somehow. But as we have seen, asynchronous DARBNs are similar to CRBNs, because both are deterministic, as opposed to ARBNs. Of course, while modelling, the choice for deterministic or non-deterministic RBNs should be also justified.

Particularly, Stuart Kauffman's work (1993) was criticized (Harvey and Bossomaier, 1997) because it assumed that genetic regulatory networks were synchronic. We agree with the critic that the synchronicity was an assumption without a base, but we do not believe that genetic regulatory networks are random. They should be most probably modelled better with DARBNs, if we can model the updating periods for each gene. But one of the main results of Kauffman was that the number of attractors on CRBNs could explain why there are very few cell types in comparison with all the possible initial states of their genes. Since the number of attractors in DARBNs is very close to the ones in CRBNs (and the number in DGARBNs is much closer, since they are semi-synchronic), we can say that this particular result, still holds (perhaps with minor adjustments), since the number of attractors does not depend too much in the synchronicity of the updates, as they depend on their determinism.

## Acknowledgements

I appreciate the valuable comments and suggestions received from Inman Harvey, Ezequiel Di Paolo, Andrew Wuensche, and two anonymous referees. This work was supported in part by the Consejo Nacional de Ciencia y Tecnología (CONACYT) of México and by the School of Cognitive and Computer Sciences of the University of Sussex.